# *Climatology of extratropical atmospheric wave packets in the Northern Hemisphere*

*Federico Grazzini*

*ARPA-SIMC Emilia-Romagna, Bologna, Italy*

*Valerio Lucarini*

*University of Reading, Dept. of Meteorology, Dept. of Mathematics, Reading, UK*

*New Draft: Versione 2.0*

———————————

*Corresponding author address* : Federico Grazzini,

ARPA-SIMC Emilia-Romagna, Viale Silvani 6, 40122 Bologna, Italy

E-mail: fgrazzini@arpa.emr.it

**ABSTRACT**

Planetary and synoptic scale wave-packets represent one important component of the atmospheric large-scale circulation. These dissipative structures are able to rapidly transport eddy kinetic energy, generated locally (e.g. by baroclinic conversion), downstream along the upper tropospheric flow. The transported energy, moving faster than individual weather systems, will affect the development of the next meteorological system on the leading edge of the wave packet, creating a chain of connections between systems that can be far apart in time and space, with important implications on predictability. In this work we present an automated recognition of atmospheric wave packets which allows the extraction of all the relevant properties, such as location, duration and velocity. Behind this tool lies the need to investigate atmospheric variability in its full complexity, bridging the low-frequency steady-state approach with the storm-tracks lagrangian approach. We have applied the algorithm to the daily analysis (every 12h) from 1958-2010, building an extended climatology of waves packets with different spectral properties. We show that wave packets characteristics over Northern Hemisphere exhibit a strong seasonal dependence, both in their spectral component and in their distribution and localization. The maximum activity is reached in the cold months, from autumn to spring, with a slight weakening in mid-winter and a clear minimum of activity in summer. Preferential areas of genesis are shown to be the Western and Central-Pacific and Western-Atlantic while areas of lyses are the eastern borders of Pacific and Atlantic. We envisage possible applications of this algorithm also for predictability studies and operational activities.

# 1. Introduction

The analysis of mid-latitude tropospheric atmospheric variability is a long-standing, crucial subject of investigation at the interface between dynamical meteorology and climate dynamics (Blackmon 1976, Speranza 1983, Hoskins et al. 1985, Schneider 2006). Whereas linear and weakly nonlinear theories allow for gaining a fundamental understanding of the mechanisms responsible for the generation and dissipation of atmospheric disturbances (Pedlosky 1987, Holton 2004, Vallis 2006), they are insufficient for the purpose of describing the dynamics of atmospheric waves in the fully nonlinear, statistically equilibrated flow (Lucarini et al. 2007a). Synoptic waves, associated with longitudinal wavenumbers above 5, are generated by ordinary baroclinic conversion (Charney 1947, Eady 1949), whereas the origin of low-frequency (low wavenumber) variability is less clear. In fact, mechanisms as different as baroclinic–orographic resonance via the form drag (Benzi et al. 1986; Ruti et al. 2006), barotropic instability of the stationary waves (Simmons et al. 1983), and Rossby wave radiation from anomalous tropical convection (Hoskins and Karoly 1981) have been proposed in this direction.

The study of the properties of synoptic and planetary scale mid-latitude atmospheric waves, like propagation, breaking and their organization is crucial for understanding weather and climate. Due to the dispersive nature of the atmosphere, mid-latitude atmospheric disturbances propagate predominantly as downstream developing waves, travelling in coherent wave packets or wave trains (Chang et al., 1999), often referred as Rossby wave packets (RWPs). RWPs grow due to baroclinic conversion in the downstream side, the leading edge, and decay barotropically in the upstream side of the packet where breaking is also observed more frequently (Lee and Held, 1993). They might be originated by a variety of meteorological processes involving for example latent heat release in the mid-lower troposphere, due for example to pre-existing extratropical cyclones or organised mesoscale convective systems, flow distortion from orography or dynamical features induced by tropical convection (Szunyogh et al., 2008). In some occasions RWPs can remain coherent over many days and propagate over long distances, tele-connecting remote regions of the atmosphere. The propagation and extension of RWPs is controlled by the intensity and localization of the background PV gradient that act as a wave guide, in analogy to electromagnetic wave propagation theory (Hoskins et al.,1985). Such coherent and long RWPs are more frequently observed in the Southern Hemisphere (SH) since the presence of vast, and baroclinically unfavourable, continental areas over the Northern Hemisphere(NH) makes them generally shorter and more difficult to track (Chang, 2001). Generally speaking, wave trains represent the large-scale carrier or precursor of "meteorological activity", accounting, with their propagation, for a large part of the observed synoptic-scale and lower frequency variability of the mid-latitude atmospheric circulation. Recently it has been shown that wave packets are also important for the shaping of the mean/zonal flow interaction (Chang 2005, Woollings et al. 2008). Several cases of severe weather triggered by RWPs are discussed in literature, few examples are: explosive cyclogenesis (Bosart et al., 1996), blockings (Renwick and Revell, 1999) and heavy precipitation over the Alpine area (Grazzini 2007, Martius et al. 2008). The possibility to automatically track RWPs, is therefore of great importance not only for research purposes but also for operational and long range forecast as stated in the Thorpex Science Plan: "The skilful prediction of [Rossby] wave-train activity is often a requisite for forecasting the synoptic-scale setting within which smaller-scale, high-impact weather events evolve" (Shapiro, 2004). For skilful NWP and climate simulations, it is therefore of primary importance that meteorological models are able to realistically reproduce the correct triggering and propagation of wave trains.

For these reasons we have developed an automated tracking algorithm of atmospheric wave packets, based on the computation of the planetary wave envelope. This approach aims to bring together apparently irreconcilable strategies of investigations of atmospheric variability such as spectral analysis of the waves proposed by Hayashi (1971, 1979), Fraedrich and Bottger (1978) and recently revived by Dell'Aquila et al. (2005) and Lucarini et al. (2007b), and the tracking of specific, objectively identified "particle-like" features, such as cyclones (Hoskins and Hodge, 2002). Our technique aims to improve on previous attempt of wave-packet tracking (see section2 for details) mainly through a more functional envelope calculation and an ad hoc developed tracking procedure. We are able to directly compute very relevant statistical properties of the wave trains as a function of their dominant wavelengths, such as mean group velocity, spatial and temporal extension, and initial and final longitude. Here we present statistical properties derived from the application of the tracking algorithm to the ECMWF ERA40 reanalysis in the period 1958-2001 and from ECMWF operational analyses for the period 2002-2010, thus covering the full period 1958-2010. We first apply our algorithm analysing the dispersive properties of RWPs composed by waves with different spectral component, ranging from zonal wavenumber 2 up to wavenumber 10. We then compare and discuss the properties of two sets of RWPs: one composed by planetary waves centred around K=3 (zonal wave number 2_4); the second composed by shorter synoptic waves centred around K=6 (5_7). In addition we perform also a statistic on RWPs centred around K=6 but with a much wider spectral band (from K=3 to K=9). The details of the algorithm and the data used will be described in section 2. In section 3 we show the results, while in section 4 we discuss implications and possible relations with other meteorological processes. Finally in section 5 we present the summary and an outlook for future work.

## 2. Wave packets tracking algorithm

The meteorological dataset consists of 12h intervals of the v-component wind at 250 hPa, taken from the ECMWF ERA40 reanalysis in the period 1958-2001 and from ECMWF operational analyses for the period 2002-2010. The v-component wind field is then interpolated onto a 2.5° x 2.5° degree regular grid resolution over the North Hemisphere. In order to extract the wave packets envelope from the v-component wind field, upon which the tracking procedure is based, we use a Hilbert transform filtering technique as implemented and described by Zimin et al. (2003). In his paper he demonstrated that the Hilbert transform technique is superior to complex demodulation, widely used in previous attempt to track wave packets (see for example Chang and Yu 1999 or Chang 2001), which is quite sensitive to the choice of the carrier wavenumber for demodulation, not known a priori. In addition, the Hilbert filtering gives the possibility to easily investigate the spectral properties of RWPs.

The technique proposed by Zimin et al. (2003) is briefly recapitulated. Given a scalar atmospheric field $\phi(s)$, with a defined value at each grid point (s) along a latitude circle, the wave packets envelope is extracted in the following way: First the Fourier transform of $\phi(s)$ is computed as

$$\hat{\phi}(k) = \frac{1}{2N+1} \sum_{j=-N}^{N} \phi(j\delta) e^{-2\pi i [kj/(2N+1)]} \qquad (1)$$

where $s = j\delta$, with $\delta$ being the grid step and $j = -N,...,0,...,N$ the grid points. The integer $k = -N,....,N$ denotes the longitudinal wavenumber. Then, the function $\hat{\phi}(k)$ is multiplied by a filter function $f(k)$ that removes all wavenumber components except for a band between positive wavenumbers $k_{min}$ and $k_{max}$. Finally the wave packets envelope is computed as the module of the inverse Fourier transform of $2f(k)\hat{\phi}(k)$.

## 2.1 Tracking procedure

Here follow a description of the main steps of the tracking procedure:

1) For each time frame (every 12h) the envelope of the v-component wind is computed as described above for each latitude.

2) An area-weighted latitudinal average of the envelope field is computed over the belt [35N-65N]. For each time frame, we find the longitude with the maximum of the envelope.

3) In order to select only significant RWPs, we defined an intensity threshold below which a new envelope maximum is discarded. For each longitude, the threshold is computed as the 7 days running mean of the average envelope over the latitudinal belt, plus an offset chosen to be 20% of the average envelope. This choice has been taken after several trials in order to retain only significant wave packets and at the same time avoiding to reduce excessively the population of the statistic. This choice was also supported by a visual match over test periods of few months, between the Hovmöller diagram (Hovmöller,1949) of the envelope and the RWPs found by the tracking algorithm (see section 2.2 and Fig.1). Note that, since the choice of the threshold is relative to the background state this has the advantage to retain also weaker summer events that would otherwise being discarded if using a fixed threshold.

4) Given an arbitrary time interval, with no pre-existing wave packets, the envelope maximum on the latitudinal belt will be the first point of a new wave packet.

5) If a point is found, in the following time steps signatures of the persistence of the wave packet are sought. The longitudinal interval over which an above-threshold point is considered as representative of the evolved wave packet found at point (4) depends of the previously estimated speed of the RWPs ± 10 [lon°/day]. The average speed of the wave packet is computed at each time step from the difference between the previous and actual position of the peak. We have introduced a smoothness criteria based upon a Gaussian weighting which favors values located at the center of the spatial interval in order to avoid abrupt changes of the RWPs speed from one temporal interval to the next, thus following the strategy usually applied in feature tracking (Hodges, 1995). This procedure repeats forward in time until the end of the packet is found.

6) The same procedure described at point 5) is then applied, starting from the same initial point, backward in time to search the beginning of the wave packet

7) The wave packet is considered dissipated when there are no points above threshold in the considered interval, both forward and backward in time. If the computed RWP lasts more than 3 days then all the longitudes and envelope intensities are saved and a new search for a new RWPs is started. A final check to avoid overlapping packets is performed, in case eliminating the one with the shortest duration.

**2.2 Postprocessing and visualization tools**

In order to qualitatively visualise its accuracy and to optimally tune the algorithm, we have designed an Hovmöller type plot where in addition to the observed v-component wind, customarily used for detecting wave trains, we display also the envelope, thus allowing a direct comparison with the tracking procedure. In Fig. 1 we show a recent example referring to a typical autumn situation with a clear propagation of a series of RWPs.

The grey shaded contours represent the latitudinal average of the envelope (over 35-65N). Single wave peaks are recognisable as streaks of higher envelope (darker grey). Solid and dashed contours represent the latitudinally averaged v-component wind at 250 hPa. Straight black lines represent wave packets automatically identified by the algorithm during this period. As can seen in this example, several RWPs are detected in this example, discriminating well periods with high RWPs activity from quiescent periods. Two majors RWPs, propagating from east Asia/western Pacific up to Europe, are detected lasting respectively 13 days (starting the 13/10 00UTC) and 8 days (starting 31/10 12UTC). As shown in this example, the algorithm performs very well in detecting initial and final longitude and duration of "isolated" and zonally travelling long lasting packets. Their integrity gets more uncertain in case the envelope is lingering over the same region at low amplitudes, like between the 25/10 and 31/10 in the example of Fig. 1, where probably a synoptic meteorologist would draw a unique longer wave packet instead of 2 as detected by the tracking procedure. This might occur when RWPs interact with persistent stationary waves which distort the waveguide or, as it is shown here, due to the interaction with other RWPs with different wavenumber and speed characteristics. A low wavenumber RWP (inferred by its slower group velocity) was emanated in fact in that area, and successively crossing the one started the 31/10 at 12UTC. In spite of the limitations here discussed, mainly attributable at the latitudinal averaging, we believe that the statistics obtained could be taken as representative and meaningful especially for the initial and final longitudes, while the explained difficulty in maintaining the correct integrity might introduce a slight underestimation concerning duration and spatial extension.

### 3. Structure and climatology of RWPs

3.1 Spectral dependence

A wave packet can be thought as composed by an infinite set of sinusoidal waves of different wavenumbers, which interfere constructively only over a small region of space, and destructively elsewhere. RWPs are dispersive wave packet in which the envelope shape is changing while propagating (Pedloski 1987, section 3.24). Hakim (2003) has tried to assess the theoretical predictions of previous studies (for example Swanson and Pierrehumbert 1994 among others) analysing the structure and the dispersion relation of developing wave packets over North Pacific and Atlantic. He showed for example how different part of wave packets, the upstream part, the packet peak, and the leading edge have different wave-numbers, group velocity and amplitudes. In the first part of this section we want to test the validity of our results against former works studying the properties of wave packets as a function of their dominant spectral component. We take full advantage of the Hilbert technique proposed by Zimin et. al (2003) that easily allow a spectral filtering on the envelope computation and therefore gives the possibility to isolate packets formed by waves belonging to a restricted zonal wavenumber interval. We have processed the atmospheric data with spectral filters retaining wavenumbers [K-1, K, K+1] with K ranging from 2 to 9, thus covering the range from very long planetary waves to synoptic waves associated with baroclinic disturbances. Figure 2 shows the distribution of number of RWPs found in each season (periods 1958-2010) as a function of the zonal wave-band used. From this figure it is evident a shift of the frequency and the distribution with season. From autumn to winter RWPs are dominated by waves with lower zonal wavenumbers, while from spring to summer there is a general decrease in the total number, and a shift towards synoptic Rossby waves with higher wavenumbers. The highest population for the different seasons are : for MAM k=6 with 704 cases, JJA k=7 with 509 cases, SON k=6 with 763 cases and DJF k=4 with 759 cases). Another important property to analyse is the dispersion relation and duration. Since the propagation speed of wave packets centered about a wavenumber K corresponds by definition to the group velocity of the wave disturbances evaluated at K, the tracking algorithm allows us to directly obtain the effective dispersion relation of (nonlinearly equilibrated) waves. Results are shown in Figure 3 for winter and summer months.

A stronger spectral dependence of velocity (dispersion) is evident in the cold months and low wavenumbers , with zonal wave number 2 and 3, significantly slower than shorter synoptic waves, with a small eastward group velocity (k=2). RWPs composed by shorter synoptic waves have higher velocities, on average 22 lon°/day (in winter), with a sharp transition around zonal wavenumber 4. The detected zonal group velocity are in the range of what is reported in literature based on observations (Chang and Yu,1999, Chang 2001) or based on numerical experiment (Simmons and Hoskins, 1979, Lee and Held, 1993).

Note that wave packets composed of shorter waves have an almost constant zonal speed of propagation so that in this range, in winter, the dispersive nature of waves is negligible. Such a weak dependence of the propagation speed with respect to the wavenumber can be interpreted in the context of the framework proposed by Eady (1949). In the Eady model, unstable waves fuelled by baroclinic conversion (as indeed the waves with waveneumber ≥ 5 are) have a phase velocity which is proportional to the wavenumber so that the kinematics is non-dispersive. In summer (fig. 3 panel c)), there is a tendency for more dispersive behaviour also in the higher wavenumber. The evidence of stronger dispersion in winter at low wavenumber is in agreement with results found by Lee and Held (1993) on a 7-year sample climatology over Southern Hemisphere (see their Fig.4). Panels b) and d) suggest a variable contribution from the different spectral component. In the cold months we observe (from fig.3 panels a) and b)) a general longer duration with a great contribution from wavenumbers around k=3 and k=4 where is also concentrated a large part of the variability (Dell'Aquila et al., 2005), being not unusual duration up to 6.5 days (mean of 4.8 days + 1σ). In summer the duration is shorter and a greater contribution comes from wavenumbers centered around K=6. Long coherent wave packets (lasting longer than 8 days in our definition) have been detected by the tracking procedure but they represent a minor although very interesting proportion (see table 1). The apparently short mean duration, compared to what is reported in literature (for example Chang 2001) could be attributed to different factors, like the choice of envelope threshold and hence the frequency of events detected and the number of spectral component used in the envelope calculation. In our setting the threshold for the envelope is variable being sometimes lower than the fixed threshold used by Chang (2010) to select only strong cases. In addition the duration reported in literature is mostly based on case studies or particular sub-set of long-lasting wave packet, like the one shown in table 1 and hereafter called wp_pac_eu, composed only by wave packets initiated in the Pacific and ending in Europe[1]. Here we present a broad statistics based on a less strict selection criteria. This could partly explain the differences together with a stronger sensitivity related to the spectral filtering used. One of the results of this investigation is that filtering over a wider spectral band is very beneficial for to get a more realistic duration of wave packets. Hakim (2003) shows for example that during their propagation RWPs undergoes transitions in their spectral component with distinct characteristics between wave peaks, the leading and the trailing edges. A narrow selection of the wave band could in this case bring to an artificial truncation of the wave packet due to the non correct sampling of all his parts. Table 1 summarises the sensitivity of the RWPs statistics to the waveband widening, showing that for RWPs centred around k=6, the duration increase drastically as more spectral component are included, especially those situated in the lower part of the spectra, while the spatial extension is not changing as much as duration since it is compensated by a slight decrease of group velocity. A comparison for example of the frequency of coherent RWPs lasting longer than 8 days deducted from table 1 and the same derived from Chang (2010) are similar (being around 3 per season in both) if we consider the wider spectral filtering 3_9.

Having shown that in order to define a RWP it useful indeed to retain a wide enough spectrum of wave component, we think it is also useful to investigate separately the various spectral

---

[1] *RWPs initiated in the west and central Pacific (between 100° and 200° Lon) and ending over Europe (between 340° and 40°E Lon)*

component contributing to RWPs characteristics. For the sake of brevity here we now concentrate on properties of RWPs composed by two distinct zonal wave band 2-4, representing long planetary and almost standing waves (here after RWPs 2_4), and to the zonal wave band 5-7, where most of the transient synoptic scale meteorological activity is concentrated (hereafter RWPs 5_7). For these two distinct types of RWPs we compute the probability distributions of the following distinctive features: mean group velocity, initial and final longitude of the packet, temporal duration, length of the path. Data are stratified according to winter and summer seasons, in order to take into consideration the planetary scale modulation of the forcing.

The distribution of mean group velocity is shown in Figure 4, panel a). It shows that RWPs 2_4 have a slower eastward group velocity than RWPs 5_7. In winter the distribution of RWPs 2_4 velocity presents a stronger deviation from gaussianity, with an hint to bimodality in proximity very small positive velocities, indicating a possible dependence of the statistical properties of large planetary waves respect from the speed of the jet (Ruti et al. 2006). Note that even if the phase speed of ultra-long waves could be very small or zero, group velocity of dispersive can have non zero group velocity. RWPs 5_7 do not exhibit a drastic change in the distribution of velocity from winter to summer, with exception of a slight decrease in the mean speed . Also in the spatial extensions (fig. 4, panel b)) the largest seasonal differences occur with RWPs 2_4, with (spatially) longer wave packets in winter, while for RWPs 5_7 there is no large seasonal difference. RWPs 5_7 are travelling over longer distances, peaking at 80 degrees of longitude of spatial extension and a long tail reaching up to 200°. RWPs 2_4 are travelling for much shorter distances peaking at 40 degrees of longitude of spatial extension. The density distribution of initial and final RWPs points (fig 4 panel c) and d) respectively), shows that there are distinct preferred spatial locations for the initiation and the decay of the packets, which depend on the considered season and on the spectral properties of the packet. Again, RWPs 2_4 show the largest changes seasonal dependence with the preferred initial longitudes shifting, from summer to winter, from the central Pacific(200°E) to the lee of Rockies Mountains (250°E), and the final longitude from the Rockies (240°E) to the central Atlantic (300°E). RWPs 5_7 are not showing radical changes with seasons. The area of genesis for the shortest waves is a broad region from east China to the central Atlantic peaking over central Pacific. In winter there is a general easterly shift to the east including also the east Atlantic Ocean. The central and eastern Atlantic appears to be regions also of dissipation of RWPs, with a very high density of decay of RWPs 5_7 in both seasons peaking at 300° east. It is also interesting to note that the main region of decay of RWPs 2_4 shifts from the Rockies mountains area to the central Atlantic in winter. A minor but interesting region of decay emerging in winter for RWPs 5_7 is the Mediterranean region (10°- 30°E). Another important feature to note is the very low density of genesis and decay for both spectral components, especially in winter, over central Asia (between 30° and 100°E). This supports previous studies indicating this region as not favourable for wave packet propagation as discussed by Park et al. (2010)

Concerning the temporal duration of the wave packets (not shown), we note that in summer RWPs 2_4 have a short temporal duration, around 3.5 days with no cases greater than 6 days. For winter we observe a common qualitative behaviour shared by RWPs 2_4 and RWPs 5_7, with a much longer tail than in summer, characterised by an exponential decrease of frequency of cases of longer duration.

3.2 Climatology of RWPs 3_9

In the previous section we have investigated how the RWPs characteristics are changing according to the dominant wavenumber of the waves composing the packets. Here we want to discuss and compare the properties of wave packets composed by waves within a wider waveband (from k=3 to k=9) including all the relevant spectral scales as can be deducted from figure 2. Although it is not clear on which degree wave packets having different wavenumber should be physically related here we assume that a single RWP, while travelling over long distances and undergoing baroclinically unstable region, could change significantly its dominating wavenumber as shown by Hakim (2003). As seen from figure 5 the monthly distribution of (coherent) RWPs 3_9 lasting more than 5 (8) days or longer than 100° (200°) of longitude displays a marked seasonal cycle with a minimum in summer months and a sharp increase from October towards the cold season. October, November, December, January and March-April are the months with the highest percentage of long RWPs (temporally and spatially). The frequency of RWPs remains high during the whole winter with a slight weakening in February which may reflect the midwinter suppression in Paficic storm-track discussed by Nakamura (1992) and Park et al. (2010). A secondary peak is present in March and April, while from May it starts the summer decrease. The mean velocity of RWPs 3_9 (figure 6) as expected increases in the cold months but the shape of its distribution is not changing radically with season like we have seen in figure 4 panel a), indicating a compensating contribution from the different spectral component. In winter the low wavenumber component are dominating, with slower speeds respects to high wavenumber waves (as we have discussed in the previous section), in summer medium to high wave number component dominate but there is an overall reduction of the jet speed. The analysis of spatial extension and temporal duration shows that the usage of a large waveband is very beneficial to achieve longer space-temporal wave packet. Longest duration and spatial extensions are achieved in autumn, winter and spring with little differentiation. Comparing results of figure 7 panel a) and b) with figure 4 and table 1 demonstrate that. This imply that RWPs might undergo to a wavenumber transition during their journey and a narrow filtering may avoid a correct detection of the leading edge, where usually a shift in frequency is observed (Hakim, 2003). Concerning the position of RWPs' initial longitude and final longitude (figure 7), the inclusion of more spectral components changes quite drastically the repartitions of areas of initiation and decay. The distribution of the initial longitudes is similar in Spring, Summer and Autumn, although in the equinox seasons there is a progressive shift of the main initiation area towards east, corresponding to Japan and East Chinese Sea. There are two other initiation areas in correspondence of the East Atlantic (280-300E) and a minor one on the Central-East Mediterranean (20-30E) in summer. In winter the main peak dominates and shifts towards the central Pacific (200E). Figure 7, panel d) shows less variability although a seasonal modulation it is still visible. Three main area of dissipation could be noticed respectively in order of importance, the Eastern Atlantic, the area around the Rockies mountains and the Central Mediterranean. A region with very low density of dissipation is in Asia between (100-150E).

The annual distribution of the total number of RWPs (figure 8) does not show any strong tendency with time except a period of a slight increase in the activity around 1990 and 2000. From the same figure it is possible to see that among an average of 82 events per year only 4 usually are coherent RWPs initiated in the west and central Pacific (wp_pac_eu). These long RWPs are important since they have been recognised to act as potential large-scale forcing for severe weather over Europe (Martius et al. 2008, Grazzini 2007).

## 4. Discussion and implications with other large-scale dynamical processes

As seen in the previous section, RWPs have distinct and robust characteristics, e.g. with a well defined initial and final locations, a marked seasonal modulation and a spectral dependence. A number of fundamental questions can then be posed in an attempt to explain the observed dynamical behaviour : 1) what are the main RWP triggers in relation with the marked distribution of the initial location? ; 2) what are the main mechanisms causing their decay and ; 3) Is the predictability of RWPs somehow related to their duration and spatial extension?

Concerning to question 1), many processes of the component flow might represent a source or trigger for RWPs. In the context of the PV framework, a positive/negative PV anomaly along the wave-guide could indeed represent an initial trigger for a RWP. Main atmospheric processes that could generate such anomalies are: a) an isentropic advection of stratospheric air from the polar vortex (positive PV anomaly); b) diabatic heating in the lower and middle troposphere (negative PV anomaly), induced for example by extra-tropical transition of tropical cyclones, warm conveyor belt of extratropical cyclones, organised mesoscale convective systems; c) flow distortion caused by topography or by large-scale dynamical features (negative PV anomaly, upper level ridge) induced by tropical convection. Martius et al. (2008) focussed their attention on the role of condensational heating in the NH storm-tracks, calculating the composite of diabating heating for days preceding the arrival of RWPs over Europe associated with extreme precipitation over the Alps, roughly corresponding to our subset denominated wp_pac_eu. They found a robust positive anomaly of diabatic heating in the same area where initiations of RWPs were detected (figure 4 of their paper). These areas of diabatic heating, and their seasonal variations, correlate very well with the initial location of RWPs found in our climatology. In particular it is worth to note that the maxima of diabatic heating (mostly attributable to baroclinic processes) occours in autumn over western Pacific, just in the same area where we have found a peak in the distribution of the initial location of RWPs 3_9 The heating patterns analysed by Martius et al. (2008) feature an evident shift to central Pacific and the appearance of a secondary maxima over the Atlantic in winter and spring, in agreement with what observed in our cases when looking at the initial location of RWPs 5_7. This strongly supports the importance of diabatic heating of antecedent synoptic systems for the seeding of new RWPs, in particular for the synoptic scale part.

The broad distribution of the density of initial location (panels c) fig. 4 e 7) and the strong correlation with diabatic heating distributions over western and central Pacific suggest that the forcing is dominated in these seasons by transient systems or by response to anomalous large-scale upper-level divergence produced by extensive areas of tropical convection (Hoskins, 1993). The sharp peak of density distribution of RWPs 3_9 initial location centred on 200° (central-Pacific) in winter, seems to indicate a more physiographically based response probably operated in concomitance with orographically induced quasi-stationary planetary waves (Held et al. 2002). If we compare the distribution of winter intial location of RWPs 3_9 with the initial location of the separated spectral component shown in fig. 4, we notice a remarkable shift to the east of the density peak of initial location (from 250° with for RWPs 2_4 to 200° with RWPs 3_9). This can be explained with the argument that as we introduced more spectral components we capture also the smaller scale upstream harmonics emanating by the planetary waves induced by the stationary planetary waves associated with the Rockies.

In addition, Kaspi and Schneider (2011) have recently shown how regions of strong latent heat fluxes over the oceans (like in boundaries regions between cold continents and warm currents like the Gulf Stream and Kuroshio) could generate stationary Rossby waves emanating RWPs mainly at low wavenumbers. This latter process, active on the eastern borders of the Pacific and Atlantic basins might help to explain the observed distribution RWPs initiation location, peaking over these regions.

Other factors may as well be relevant. Cassou (2008) explored the teleconnection between MJO and NAO. He gave statistically robust results showing that the massive upper level divergent flow, associated with the convection of MJO in phase 3/4 (over Indonesia, where it reaches its maximum in its 40 days cycle), acts as a Rossby wave vorticity source, inducing a downstream quasi-stationary wave packet in the extratropical NH circulation. This RWP has a low zonal wavenumber (2-3), stretches from western-Pacific to Europe and moves very slowly eastward. This also might be processes acting as source of low wavenumber RWPs, although more variable in time.

The duration of single RWPs (in space and time), is generally controlled by the strength and position of the background PV gradient acting a s a waveguide. Stronger PV gradients produce a higher zonal trasmissivity and greater coherence of the waves, weaker gradients tend to absorb the incoming waves (Martius et al. 2008). Chang and Yu (1999) have shown that for wintertime circulation (their figure 7 e)) the regions of tighter gradients of PV on the isentropic surface of 350 K are located downstream the main land masses, over extratropical western-Pacific and U.S. east coast, while regions of diffluence (absorbtion of RWPs) are located upstream NH continents, over western U.S and in particular over eastern Atlantic/western Europe. A very weak gradient is present over continental Asia. This picture is fully consistent with our climatology of preferred initial and final location showing that regions of preferred initiation are regions favourable for RWPs transmissions and triggering (as discussed above), while area where indicated as final longitudes of RWPs are indeed areas not favourable for RWPs transmissions as it clearly stands out, from panels d) of figure 4 and 7, to be the central-east Atlantic. The weak PV gradient over central Asia north of the Himalaya mountains supports also the lack of activity we have found over the continent. Actually, most of the remaining waves over this region transit slightly south of 35°N (Chang and Yu 1999).

It is worth discussing the possible effects on predictability in case of presence of especially long lasting coherent RWPs. This physical connection operated by the downstream development at very large spatial-temporal scales, sometimes spanning half of the hemisphere and lasting many days, might induce to think as the atmosphere would transit in a phase state with "reduced degrees of freedom", with potentially higher predictability. Although this, by default, does not guarantee better predictive skill if the prediction of the physical processes responsible for example of the initiation and propagation of the Rossby wave packet are not captured correctly, this might still represent a tangible increase of predictability due to the boundary confinement of errors inside the waveguide. Empirical evidence, shown by Grazzini (2007, 2009), suggests that, in a state of the art NWP model, the representation of the dynamical processes along the waveguides is sufficiently accurate to ensure higher than average predictive skill of some of the downstream meteorological events part of propagation RWPs.

Another indirect piece of evidence of increased predictability comes from a recent study of Vitart and Molteni (2009) in which they show a clear increase in predictive skill in week2/week3 forecast once and MJO in phase 2-4 triggers a RWP in the initial conditions.

Given these arguments and also as mentioned by Lee and Held (1993), due to the coherence of the packet respect to its chaotic internal dynamics, the packet envelope should be more predictable than individual weather systems. To exploit the envelope predictability we envisage a potential use of the RWPs tracking algorithm and his visualising tool (as shown in Fig.1) in operational applications, especially for long range ensemble systems (monthly and seasonal forecast). If the predictability of individual waves becomes at most very uncertain at such long ranges, it might be useful to concentrate on more general properties the atmosphere like for example its capacity to transmit wave packets originated from upstream region. This tool, applied to the model output of long-range forecast suites, might be used to anticipate periods with high RWPs trasmissivity from quiescent periods characterized by low large-scale meteorological activity and weak meridional exchanges.

## 5. Summary and outlook

We can summarise the results as follows:

- Wave packets in the NH mid-latitudes (35N-65N) exhibit a strong seasonal cycle with a frequency maximum in the cold months, from October to April. Interestingly October and November months are the months with the overall greatest activity in general and also with a maximum of frequency of coherent and longer wave packet, often directly connecting west-Pacific and Europe (wp_pac_eu). Summer months have a minimum of activity. No significant trend has been observed in the yearly number of RWPs in the analysed period (1958-2010).

- We have investigated the spectral properties of wave packet showing that a stronger dispersion is more evident in winter and at low zonal wavenumeber, with $k \leq 4$. In winter and autumn the packets peak are dominated by waves with zonal wavenumber mostly ranging from $k=4$ to $k=6$, while from spring to summer there is a progressive shift towards $k=6/7$ . RWPs are generally slower in summer months and at low wavenumbers.

- Coherent and long events (packets lasting more than 8 days) are relatively rare, with packets directly connecting the western or central Pacific with Europe in the order of 4 per year. The bulk of the distribution is composed by events with duration comprised between 3 days and 6 days according to the season. A greater duration is observed at lower zonal wavenumber in winter and for zonal wavenumber $k=6$ in summer. Widening the filtering wave-band is very beneficial for the detection of long RWPs, suggesting a physical relation of processes acting at different scales, like for example in the hypothesis of synoptic scale wave packets radiating from quasi-stationary planetary waves.

- There are distinct area of genesis and decay in relation with a given spectral component and a season of the year pointing to particular atmospheric processes responsible for decay and initiation. Preferred areas of genesis are the Pacific and western Atlantic. The western Pacific tends to be a predominant area of genesis in autumn and spring while in summer most of the wave packets originates in the central Pacific. The maximum of genesis over the Atlantic is recorded in spring. The decay areas are the Mediterraean area, in the Rockies mountains region and the central and eastern Atlantic which show the highest density of decay in all seasons, associated with a diffluent gradient of PV at the tropopause level not favourable for RWPs trasmissivity. Over continental Asia, in the band of latitude (35N-65N), RWPs initiation/decay it is very low reaching a minimum.

We have presented a practical complementary approach to the study of atmospheric wave packet, aiming at consider atmospheric processes in their complexity and interdependence and not a sum of single entities. Through an automated tracking procedure we have produced a systematic climatology of NH atmospheric wave packets over the last 53 years further reinforcing the body of observational studies pointing to important differences respect to what predicted by linear theory which allows for example a symmetric upstream and downstream expansion of the wave packet. In addition this climatology can be used as powerful tool to diagnose model errors related to systematic errors of wave packets propagation.

In the future, relaying on a refined version of the envelope computation developed by Zimin et al. (2006), we would like to continue the study of the properties of atmospheric wave packets closely investigating the triggering of RWPs and further exploring their potential predictability at longer ranges. Other future studies involve the evaluation of the climatology of the mid-latitude wave packets in the suite of climate models included in the World Climate Research Programme's (WCRP's) Coupled Model Intercomparison Project phase 3 (CMIP3) multi-model dataset, in order to further characterise, following Lucarini et al. (2007b), their coherence in the representation of extratropical atmospheric variability in response to changing climatic conditions.

*Acknowledgments*

We would like to thank Istvan Szunyogh and collaborators of the department of atmospheric science of the University of Maryland (Washington D.C.) for useful suggestions and for providing a refined version of the envelope computation. The first author wish to thank U.S. Office of Naval Research Global, Visiting Scientist Program (*FY08 4015)*, for providing financial support for the visit at University of Maryland.

a)

| 5_7 | n | Cg [lon°/day] | > 5d | > 8d | > 100°Lon | > 200°Lon |
|---|---|---|---|---|---|---|
| SON | 763 | 21.3 ± 6.4 | 28% | 3% | 33% | 4% |
| DJF | 657 | 21.8 ± 6.1 | 22% | 1% | 31% | 3% |
| MAM | 704 | 20.1 ± 6.2 | 27% | 3% | 28% | 2% |
| JJA | 489 | 19.1 ± 5.1 | 22% | 2% | 20% | 1% |
| *wp_pac_eu* | 57 | 28.5 ± 4.6 | 100% | 21% | 100% | 44% |

b)

| 4_8 | n | Cg [lon°/day] | > 5d | > 8d | > 100°Lon | > 200°Lon |
|---|---|---|---|---|---|---|
| SON | 1020 | 20.8 ± 4.7 | 41% | 9% | 41% | 6% |
| DJF | 988 | 21.1 ± 5.0 | 40% | 8% | 41% | 6% |
| MAM | 1026 | 19.3 ± 4.6 | 36% | 6% | 32% | 3% |
| JJA | 882 | 18.6 ± 4.5 | 30% | 5% | 23% | 2% |
| *wp_pac_eu* | 191 | 25.3 ± 4.2 | 100% | 52% | 100% | 48% |

c)

| 3_9 | n | Cg [lon°/day] | > 5d | > 8d | > 100°Lon | > 200°Lon |
|---|---|---|---|---|---|---|
| SON | 1109 | 19.7 ± 5.0 | 47% | 13% | 43% | 7% |
| DJF | 1088 | 19.6 ± 5.0 | 51% | 13% | 45% | 6% |
| MAM | 1109 | 18.0 ± 4.7 | 46% | 13% | 36% | 4% |
| JJA | 1019 | 17.2 ± 4.6 | 35% | 7% | 24% | 1% |
| *wp_pac_eu* | 209 | 23.4 ± 4.0 | 100% | 70% | 100% | 44% |

Table 1: panels a), b), c) are summarising respectively the main characteristics of RWPs composed by waves centred around K=6 with increasing zonal waveband widths (5_7, 4_8, 3_9) of the spectral filtering. In addition a subset of long wave packet, spanning from the Pacific to Europe, is also included for comparison (wp_pac_eu). Total number of events, mean group velocity, frequency of RWPs lasting more than 5 days and 8 days, frequency of RWPs longer than 100° Lon and 200° Lon are respectively shown in each table for each season.

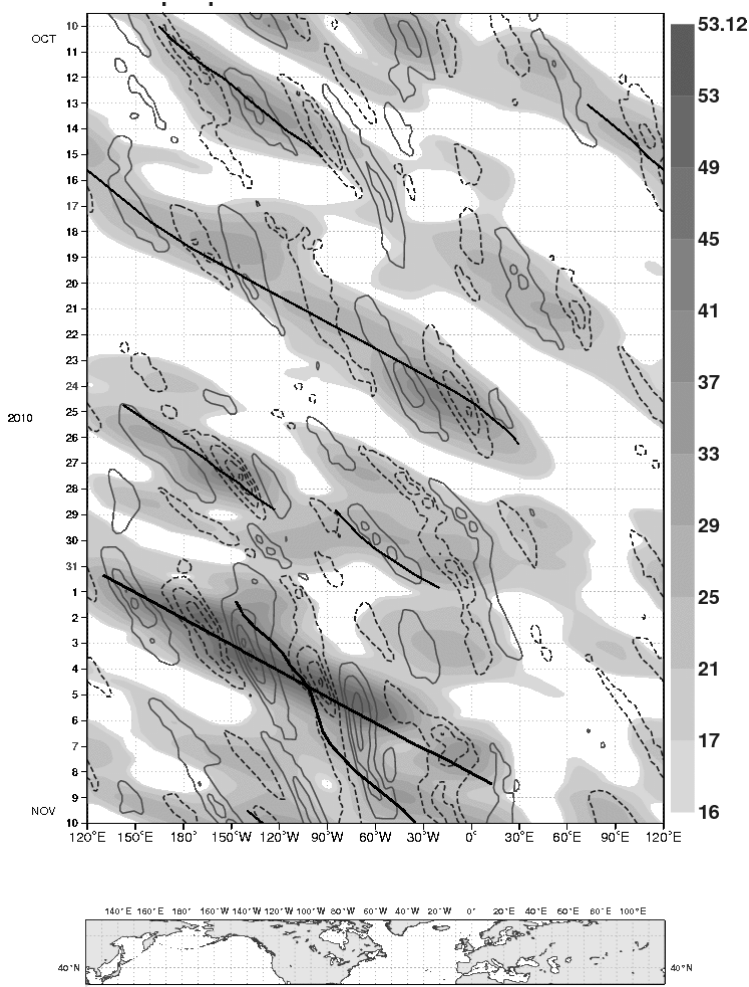

Figure 1: Hovmöller representation of RWPs propagation. Time is on the y-axis, starting at 10/10/2010 00 UTC (top) and ending at 10/11/2010 00 UTC (bottom). On the x-axis, the full belt of North Hemispheric longitudes is represented, starting from 120E . The little map below helps to associate longitudes with the corresponding geographic location. The greyscale shaded contours represent the latitudinal averaged envelope (m s$^{-1}$)over 35-65N. Visually, single wave packets are recognisable as propagation of streaks of higher envelope (dark grey). Solid and dashed contours

represent the v-component wind at 250 hPa (m s$^{-1}$, plotted every 15 m s$^{-1}$.solid contours are positive values (southerly flow), dashed contours are negative values (northerly flow). The black straight lines are RWPs detected by the algorithm in the period. In this example,the spectral band used to compute the envelope is between k = 3 and k = 9.

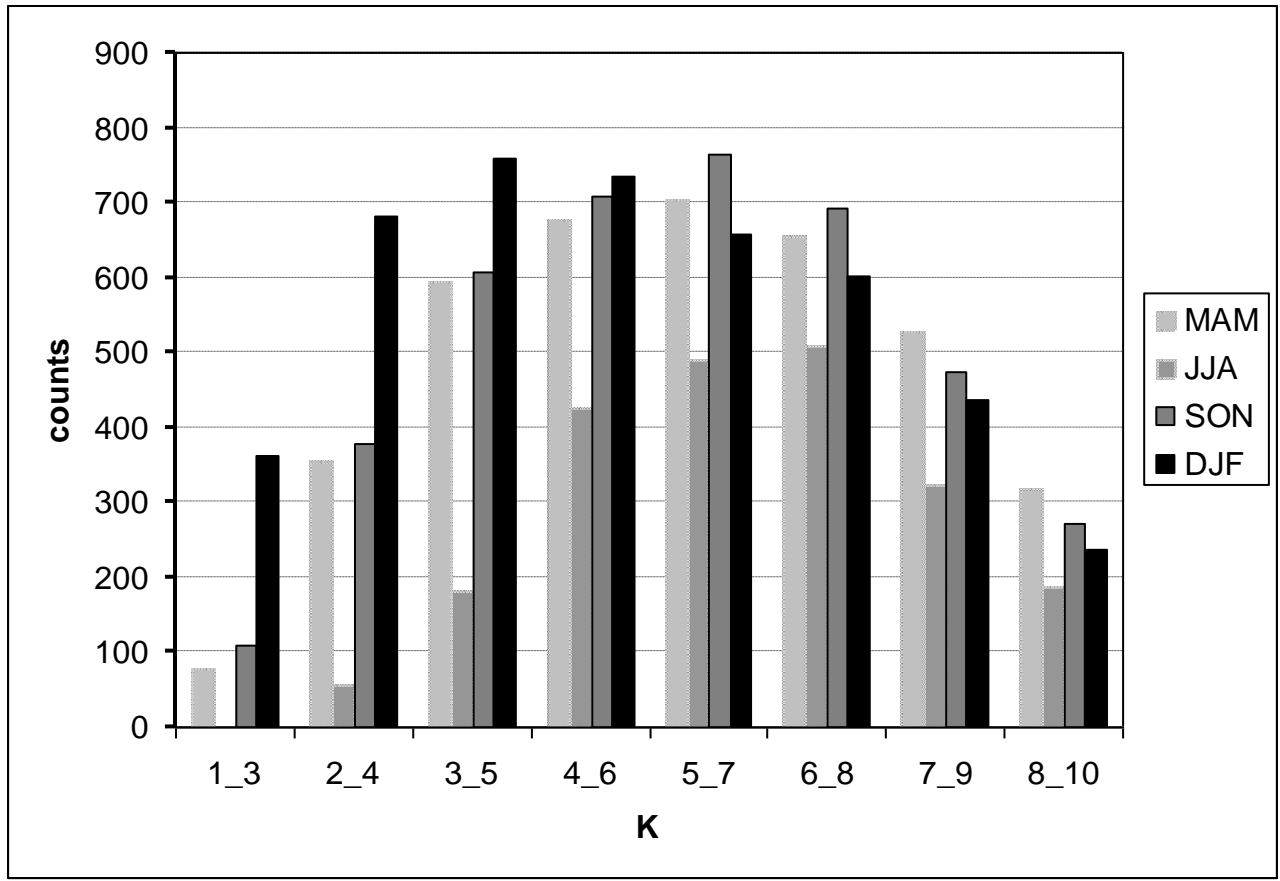

Figure 2: Distribution of number of RWPs find in each season as a function of the waveband (period 1958-2010). Note how from autumn to winter RWPs are dominated by waves with low zonal wavenumbers (central wave number K=4 is the most frequent in winter with 759 events) while from spring to summer there is a general decrease in the total number and RWPs are composed by waves with higher wavenumber (k=7 have the highest counts in summer with 509 events).

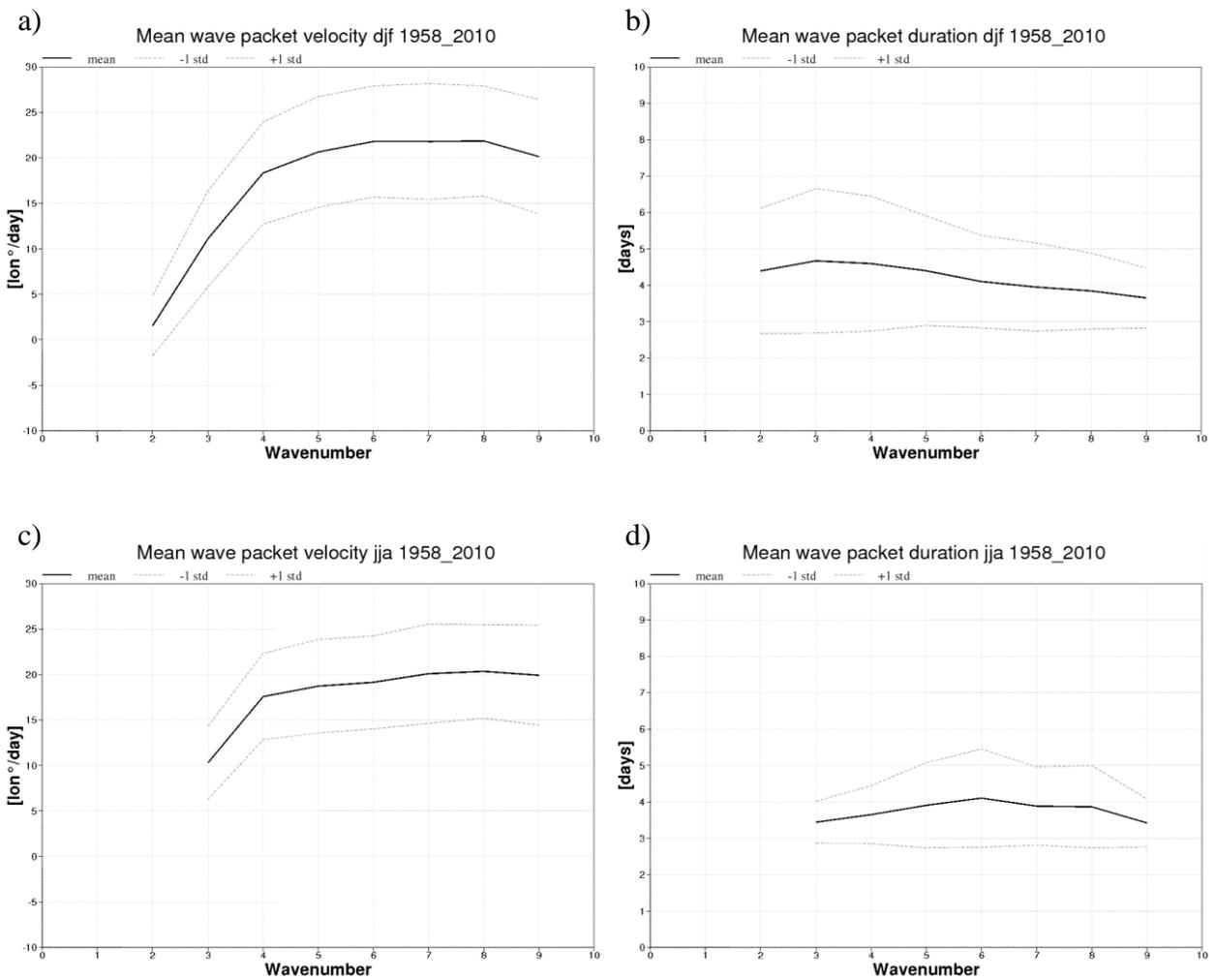

Figure 3: Panel a) and c) show the dependence of mean wave packet velocity [lon°/day], (1°lon/day~1m/s at mid-latitudes) on the central zonal wave number K. Panel b) and d) shows the average duration in time of the wave packet versus K. The intervals of ±1σ are also shown. In the top row the statistics computed over winter seasons (djf) is reported while in the bottom row it is shown the summer statistics (jja). Period 1958-2010.

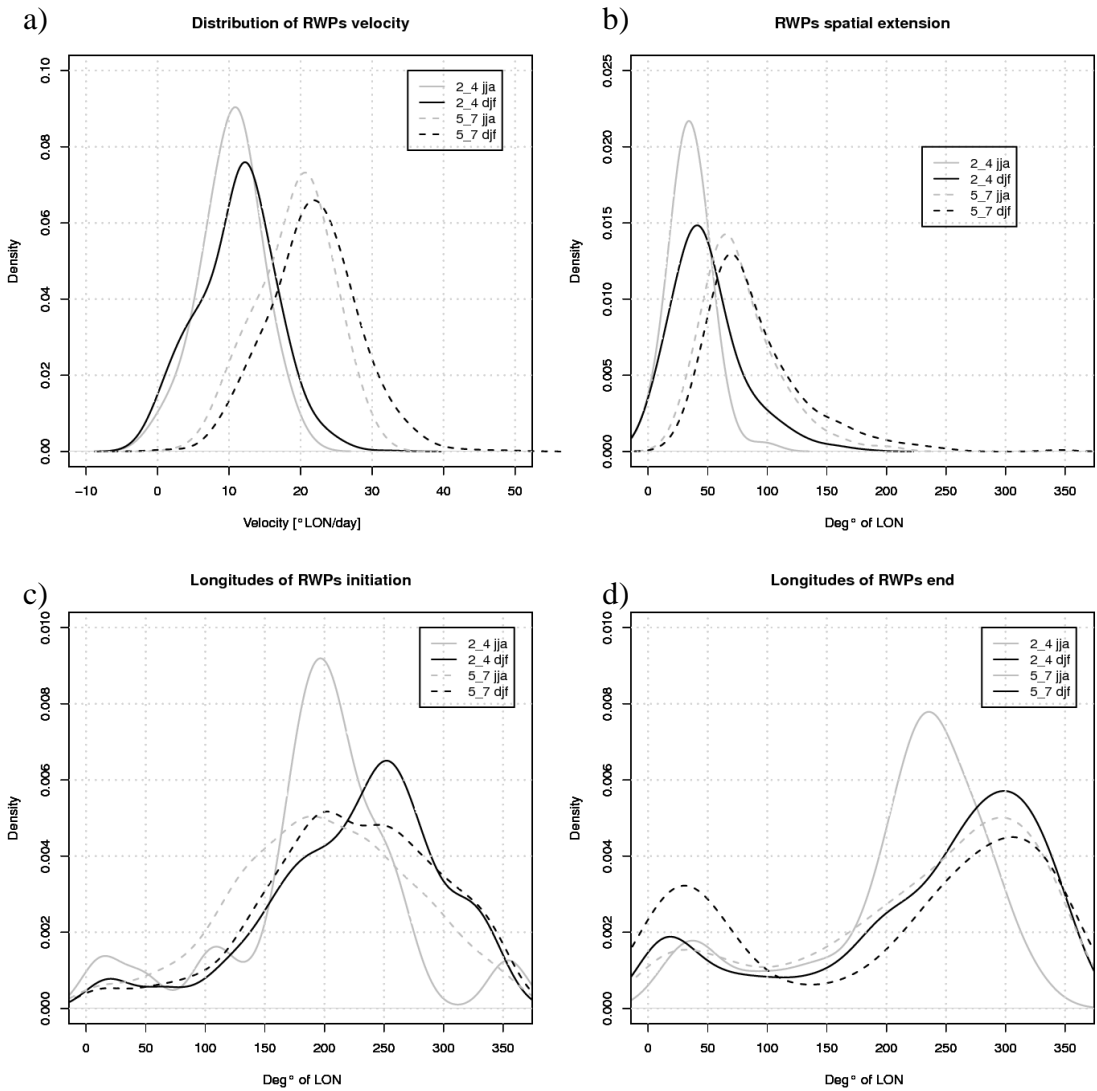

Figure 4: Comparison of density distribution of characteristic parameter of RWPs composed by different types of waves: solid lines refer to RWPs 2_4 (planetary waves), dashed lines refer to RWPs 5_7 (synoptic waves). Grey lines refer to summer months, black lines refer to winter months (period 1958-2010). Panel a) shows the mean group velocity, b) shows the distribution of the spatial extension, c) and d) are showing respectively the distribution of the initial and final longitudes of RWPs

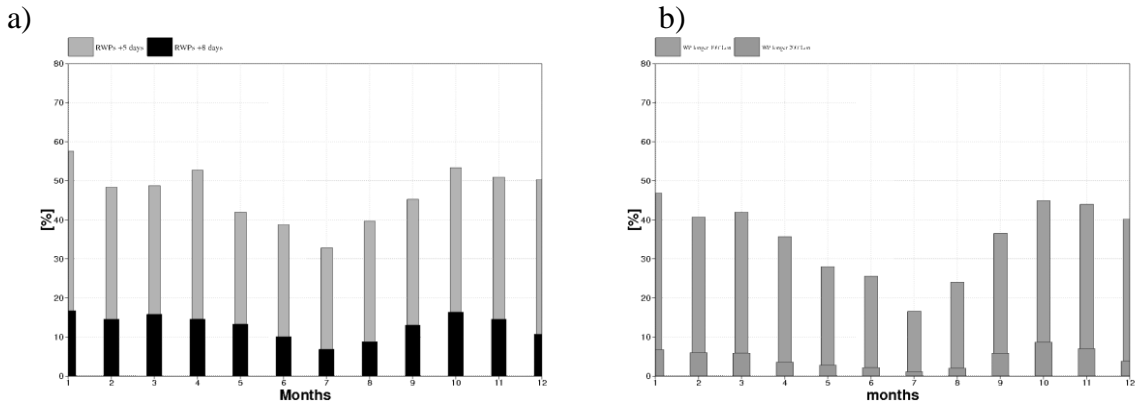

Figure 5: Panel a) shows the monthly distribution of percentage of RWPs lasting longer than 5 days (grey bars) and 8 days (black bars). Panel b) shows the monthly distribution of RWPs spanning distances longer than 100° of longitude. (grey bars), and longer than 200° of longitude (larger bars)

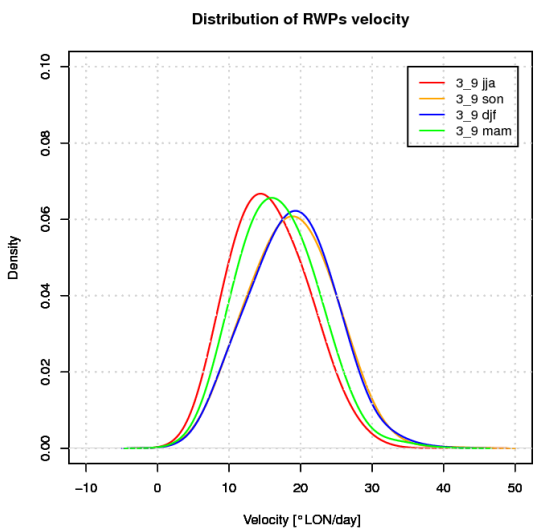

Figure 6: Density distributions of the mean velocity of RWPs 3_9 according with seasons (period: 1958-2010).

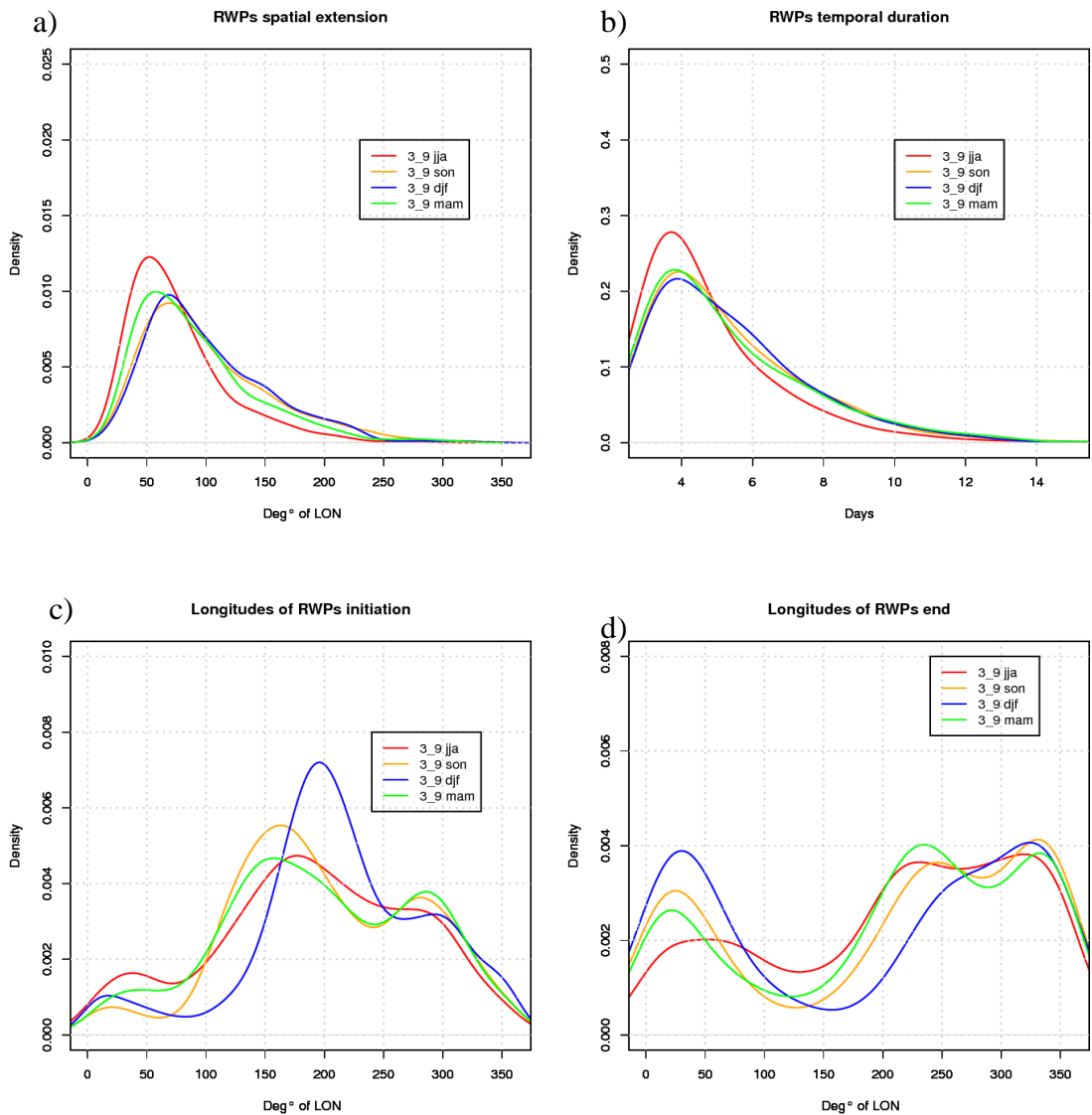

Figure 7: Density distributions of the main characteristics of RWPs composed by waves in the zonal wave band 3-9 as function of the season of the year (period: 1958-2010). Panel a) shows the mean group velocity, b) shows the distribution of the spatial extension, c) and d) are showing respectively the distribution of the initial and final longitudes of RWPs

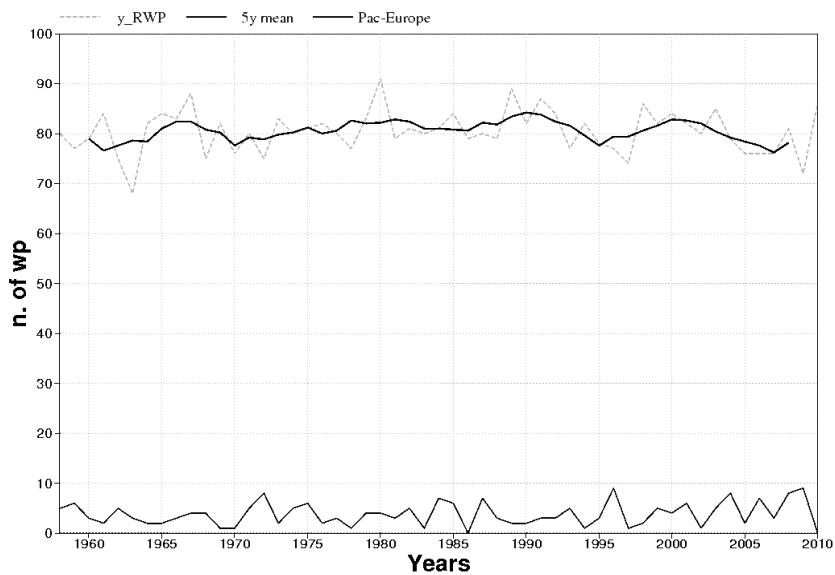

Figure 8: Distribution of the annual number of RWPs (3_9). The dash grey line indicates the yearly counts of RWPs (mean: 82, standard deviation: 4) [counts/year]. The thick black line shows a 5 year centred running mean. The thin black line at the bottom shows the yearly number of RWPs initiated in the west and central Pacific (between 100° and 200° Lon) and ending over Europe (between 340° and 40°E Lon). Such long wave packets (wp_pac_eu) are very rare, about 4 every year.